\documentclass[aps,prb,twocolumn,superscriptaddress,showpacs]{revtex4}
\usepackage{graphicx}
\begin{document}

\title{Contribution of the second Landau level to the exchange energy of the 
three-dimensional electron gas in a high magnetic field}
\author{J. M. Morbec}
\affiliation{Departamento de F\'{\i}sica e Inform\'atica,
Instituto de F\'{\i}sica de S\~ao Carlos\\
Universidade de S\~ao Paulo,
Caixa Postal 369, S\~ao Carlos, 13560-970 SP, Brazil}
\author{K. Capelle}
\email{capelle@if.sc.usp.br}
\affiliation{Departamento de F\'{\i}sica e Inform\'atica,
Instituto de F\'{\i}sica de S\~ao Carlos\\
Universidade de S\~ao Paulo,
Caixa Postal 369, S\~ao Carlos, 13560-970 SP, Brazil}
\affiliation{Theoretische Physik, Freie Universit\"at Berlin, D-14195 Berlin,
Germany}

\date{\today}

\begin{abstract}
We derive a closed analytical expression for the exchange energy of the
three-dimensional interacting electron gas in strong magnetic fields, which 
goes beyond the quantum limit ($L=0$) by explicitly including the effect of 
the second, $L=1$, Landau level and arbitrary spin polarization. The inclusion 
of the $L=1$ level brings the fields to which the formula applies closer to 
the laboratory range, as compared to previous expressions, valid only for 
$L=0$ and complete spin polarization. We identify, and explain, two distinct 
regimes, separated by a critical density $n_c$. Below $n_c$, the per-particle
exchange energy is lowered by the contribution of $L=1$, whereas above $n_c$ 
it is increased. As special cases of our general equation we recover 
various known, more limited, results for higher fields, and identify and 
correct a few inconsistencies in some of these earlier expressions. 
\end{abstract}


% 71.10.Ca Electron gas, Fermi gas
% 71.15.Mb Density functional theory, local density approximation, gradient
% and other corrections
% 71.70.Di Landau levels
% 97.60.Jd Neutron stars

\maketitle

\newcommand{\be}{\begin{equation}}
\newcommand{\ee}{\end{equation}}
\newcommand{\bea}{\begin{eqnarray}}
\newcommand{\eea}{\end{eqnarray}}
\newcommand{\bi}{\bibitem}

\renewcommand{\r}{({\bf r})}
\newcommand{\rp}{({\bf r'})}

\newcommand{\ua}{\uparrow}
\newcommand{\da}{\downarrow}
\newcommand{\la}{\langle}
\newcommand{\ra}{\rangle}
\newcommand{\dg}{\dagger}

\section{Introduction}
\label{intro}

The calculation of the exchange energy of an interacting Fermi gas in high 
magnetic 
fields is a fundamental problem of many-body physics, with applications in
fields as diverse as semiconductor physics,\cite{lai,landwehr1,landwehr2} 
astrophysics,\cite{lai,ruder,schmelcher} atomic and molecular 
physics\cite{lai,ruder,schmelcher} and density-functional 
theory.\cite{vr1,vr2,harris1,harris2} 

In a seminal 1971 paper\cite{dg71} Danz and Glasser (hereafter DG) 
calculated the exchange energy, $e_x$, of a three-dimensional electron gas
in high magnetic fields, by means of Green's function techniques. A key result
are analytical expressions for the dependence of $e_x$ on density $n$ and 
magnetic field $B$, valid if the electrons are fully spin polarized and 
occupy only the spin-down sublevel of the lowest Landau level. In an equally 
important 1974 paper Banerjee, Constantinescu and Rehak\cite{bcr74} 
(hereafter BCR) also calculated this exchange energy, 
and obtained a result that looks very similar to that of DG. The calculations 
of DG and BCR provided the background for a large body of later work on the 
exchange and correlation energy of the electron gas in high magnetic 
fields\cite{horing,yonei,fushiki,fushiki2,lai,ortner,skudlarski,takadagoto} 
and are also frequently quoted as input for the local-density approximation to 
current-density-functional theory in strong magnetic 
fields.\cite{skudlarski,rasolt,daragosh,jones,roadprl}

A major limitation of these early calculations is their restriction to
complete spin polarization and to the lowest Landau level, which in three 
dimensions requires either magnetic fields that are beyond what is 
currently achievable in the laboratory, or restriction to low-density 
low-effective-mass systems. Here we extend the DG many-body calculations
to the case of arbitrary spin polarization, and include the contribution 
of the second, $L=1$, Landau level. As a consequence, the range of magnetic
fields to which the resulting expression applies is extended towards weaker 
fields, as compared to earlier expressions.

Our formula reveals rather complex behaviour of the exchange energy, once 
higher Landau-levels are included: As a function of the density $n$, $e_x$ 
first drops with a discontinuous derivative at density $n_d$, corresponding 
to the onset of occupation of $L=1$, and then passes through two regimes, 
separated by a critical density $n_c$. Below $n_c$, the per-particle exchange 
energy is lowered (in modulus) by occupation of the $L=1$ level, whereas above 
$n_c$ it is increased (in modulus). The crossover between
both regimes corresponds to a local minimum of $e_x(n)$ at $n_c$.
The physics of the drop and of both regimes can be understood in terms of
the Landau-level structure. For currently achievable fields, $n_c$ falls 
into the metallic density regime, and thus should be observable.

Next, we consider various special cases of our general expression, in 
order to make contact with more restricted results previously available
in the literature, in particular those of DG and BCR. Scrutiny of these 
earlier expressions reveals a number of small inconsistencies and mistakes,
which we correct on the basis of our more general expression.

\section{Exchange energy including the second Landau level}
\label{calc}

The extension of the DG calculation to include higher Landau levels had, up to 
now, not been achieved in closed from. DG, and many other workers,\cite{bcr74,horing,yonei,fushiki,fushiki2,lai,ortner,skudlarski,takadagoto} go beyond the
quantum limit by rewriting the exchange integrals via expansion in infinite 
series, which cannot be resummed, or calculate them numerically. Neither 
approach yields analytical expressions that can be used, {\em e.g.} in the 
construction of density functionals for current-density-functional 
theory.\cite{vr1,vr2}
Motivated by the need for analytical expressions for lower fields,
including $L=1$, and by the observation of small inconsistencies in available 
results for $L=0$ (see below), we have recalculated the exchange energy along 
the same lines as in the DG calculation for $L=0$, but keept all
contributions from the $L=1$ level. The result turns out to permit a closed
analytical expression which, although lengthy, can be expressed in terms of
the same special functions and physical variables as the original DG 
formula for $L=0$. To be concise, we here just present the final result. 
More details on its derivation can be found in Appendix \ref{appA}. 

Three types of terms contribute: one, $e_x^{(0)}$, arises exclusively from 
$L=0$, another, $e_x^{(1)}$, arises from $L=1$, and a third,
$e_x^{(0,1)}$, from inter-level exchange, involving contributions from 
$L=0$ and $L=1$. As a function of the occupation numbers $n_L^{\sigma}$ of
the spin up and down sublevels of the $L=0$ and $L=1$ Landau levels,
the final result for the per-volume exchange energy can be written 
\begin{widetext}
\be
e_x(n_0^{\uparrow},n_0^{\downarrow},n_1^{\uparrow},n_1^{\downarrow},B)
=\frac{e^2}{8\pi^3}\left(\frac{m\omega_c}{\hbar}\right)^2\sum_{\sigma}\left[
e_x^{(0)}(n_0^\sigma,B)+
e_x^{(0,1)}(n_0^\sigma,n_1^\sigma,B)+
e_x^{(1)}(n_1^\sigma,B)
\right]
\label{exLone}
\ee
where
\begin{eqnarray}
e_x^{(0)}(n_0^\sigma,B)
=C+\ln{\left(\frac{8\pi^4\hbar^3}{m^3\omega_c^3}{n_0^{\sigma}}^2\right)}-\exp{\left(\frac{8\pi^4\hbar^3}{m^3\omega_c^3}{n_0^{\sigma}}^2\right)}Ei\left(-\frac{8\pi^4\hbar^3}{m^3\omega_c^3}{n_0^{\sigma}}^2\right) \nonumber\\
-\left(\frac{8\pi^4\hbar^3}{m^3\omega_c^3}{n_0^{\sigma}}^2\right)^{1/4}G_{23}^{22}\left(\frac{8\pi^4\hbar^3}{m^3\omega_c^3}{n_0^{\sigma}}^2\left|\begin{array}{l}
3/4, \, 5/4   \\
3/4, \, 3/4, \, 1/4\\
\end{array}
\right.\right),
\end{eqnarray}
\begin{eqnarray}
e_x^{(0,1)}(n_0^\sigma,n_1^\sigma,B)
=4\ln{\left(\frac{n_0^{\sigma}+n_1^{\sigma}}{n_0^{\sigma}-n_1^{\sigma}}\right)}+2\exp{\left(\frac{2\pi^4\hbar^3}{m^3\omega_c^3}\left(n_0^{\sigma}-n_1^{\sigma}\right)^2\right)}Ei\left(-\frac{2\pi^4\hbar^3}{m^3\omega_c^3}\left(n_0^{\sigma}-n_1^{\sigma}\right)^2\right)\nonumber\\
-2\exp{\left(\frac{2\pi^4\hbar^3}{m^3\omega_c^3}\left(n_0^{\sigma}+n_1^{\sigma}\right)^2\right)}Ei\left(-\frac{2\pi^4\hbar^3}{m^3\omega_c^3}\left(n_0^{\sigma}+n_1^{\sigma}\right)^2\right)  \nonumber\\
+\left(\frac{2\pi^4\hbar^3}{m^3\omega_c^3}\left(n_0^{\sigma}-n_1^{\sigma}\right)^2\right)^{1/4}G^{22}_{23}\left(\frac{2\pi^4\hbar^3}{m^3\omega_c^3}\left(n_0^{\sigma}-n_1^{\sigma}\right)^2
\left|\begin{array}{l}
3/4, \, 5/4   \\
3/4, \, 3/4, \, 1/4\\
\end{array}
\right.\right)\nonumber \\
-\left(\frac{2\pi^4\hbar^3}{m^3\omega_c^3}\left(n_0^{\sigma}+n_1^{\sigma}\right)^2\right)^{1/4}G^{22}_{23}\left(\frac{2\pi^4\hbar^3}{m^3\omega_c^3}\left(n_0^{\sigma}+n_1^{\sigma}\right)^2\left|\begin{array}{l}
3/4, \, 5/4   \\
3/4, \, 3/4, \, 1/4\\
\end{array}
\right.\right),
\end{eqnarray}
and
\begin{eqnarray}
e_x^{(1)}(n_1^\sigma,B)
=C+\ln{\left(\frac{8\pi^4\hbar^3}{m^3\omega_c^3}{n_1^{\sigma}}^2\right)}-\exp{\left(\frac{8\pi^4\hbar^3}{m^3\omega_c^3}{n_1^{\sigma}}^2\right)}Ei\left(-\frac{8\pi^4\hbar^3}{m^3\omega_c^3}{n_1^{\sigma}}^2\right)\nonumber\\
-\frac{3}{4}\left(\frac{8\pi^4\hbar^3}{m^3\omega_c^3}{n_1^{\sigma}}^2\right)^{1/4}G_{23}^{22}\left(\frac{8\pi^4\hbar^3}{m^3\omega_c^3}{n_1^{\sigma}}^2\left|\begin{array}{l}
3/4, \, 5/4   \\
3/4, \, 3/4, \, 1/4\\
\end{array}
\right. \right)\nonumber \\
+\frac{1}{2}\left(\frac{8\pi^4\hbar^3}{m^3\omega_c^3}{n_1^{\sigma}}^2\right)\exp{\left(\frac{8\pi^4\hbar^3}{m^3\omega_c^3}{n_1^{\sigma}}^2\right)}Ei\left(-\frac{8\pi^4\hbar^3}{m^3\omega_c^3}{n_1^{\sigma}}^2\right).
\label{exLone4}
\end{eqnarray}
\end{widetext}

Here $\omega_c(B) = e B/mc$ is the cyclotron frequency, $C=0.57722$ is Euler's 
constant, $G^{22}_{23}$ is the Meijer $G$ function \cite{luke} and $Ei$ is the 
exponential integral. In principle, this expression hold for arbitrary values 
of the g-factor, as long as $L\leq 1$, but we note that if 
the free-electron value $g=2$ is employed, an accidental degeneracy between the
spin-up subband of $L=1$ level and the spin-down subband of $L=2$ occurs. 
Restriction to $L\leq1$ is thus only rigorously possible if either $g<2$ (a 
common situation in semiconductors) or if the $L=1$ level is fully polarized, 
so that its spin-up subband is empty.

The condition $L\leq 1$ implies a restriction on the allowable 
values of density and magnetic field. In terms of the magnetic length
$l(B)=\sqrt{\hbar c/(eB)}$ and the density parameter 
$r_s(n)a_0=[3/(4\pi n)]^{1/3}$, where $a_0=\hbar^2/(me^2)$ is the Bohr 
radius, this restriction is conveniently written as
\be
\frac{l}{r_sa_0} < \left[\frac{4}{3\pi}\left(1+\sqrt{2}\right)\right]^{1/3}
\label{l1g2cond}
\ee
for $g=2$, and
\be
\frac{l}{r_sa_0} < \left[\frac{4}{3\pi}\left(2+\sqrt{2}\right)\right]^{1/3}
\label{l1g0cond}
\ee
for $g=0$. These conditions are derived in Appendix \ref{appB}.

Regarding the spin dependence, we note that spin up and spin down contributions
do not mix, {\em i.e.}, their contribution to $e_x$ can be evaluated
separately. Equations (\ref{exLone}) to (\ref{exLone4}) permit one to do 
this for arbitrary values of the occupation numbers $n_L^\sigma$, and thus
also for arbitray spin polarizations $m\propto (n_\ua - n_\da)$, 
where $n_\sigma=\sum_L^{occ}n_L^\sigma$.

\begin{figure}
\includegraphics[height=60mm,width=75mm,angle=0]{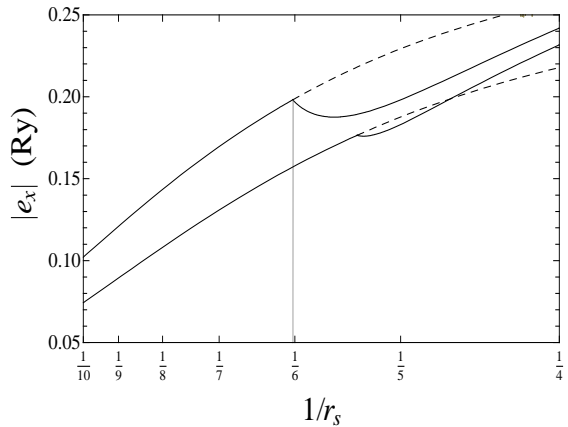}
\caption{\label{fig1}
Full curves: Exchange energy per particle, obtained from Eq.~(\ref{exLone})
by dividing by the density,
as a function of $r_s^{-1} \propto n^{1/3}$. Dashed curves: (Erroneous) 
continuation of the $L=0$ expression into the $L=1$ regime.
Upper set of two curves: $g=2$ and $B=1.4448\times 10^4 T$.
Lower set of two curves: $g=0$ and $B=1.1474\times 10^4 T$.
}
\end{figure}

As illustration of Eq.~(\ref{exLone}), Fig.~\ref{fig1} displays the 
exchange energy for a combination of densities and magnetic fields 
for which both $L=0$ and $L=1$ levels contribute, and compares it with 
the (erroneous) use of the $L=0$ expression alone in the same regime. 

Figure~\ref{fig1} reveals interesting behaviour of $e_x$ that appears only 
once $L>0$ is allowed for: Upon adding more particles (increasing the 
density), the per-particle exchange energy suddenly drops (in modulus) once 
the $L=1$ level starts to be occupied at density $n_d$, and then passes 
through two regimes. Initially, it continuous to decrease, while
for larger densities it increases (in 
modulus) up to values larger than those obtained by allocating all particles 
in $L=0$. The first regime is entered with a discontinuous derivative in the 
$e_x(n)$ curve, indicating a zero-temperature phase transition. The second 
regime is entered via a gradual crossover. Both regimes are separated by a 
critical density $n_c$, at which $e_x(n)$ goes through a local minimum.

This intricate behaviour finds its explanation in the Landau-level structure: 
Due to the differences in the spatial part of the Landau-level wave functions,
intra-level exchange integrals
are larger than inter-level integrals. As long as only $L=0$ 
contributes, the exchange energy naturally increases with the number of 
particles. Once particles are allocated in the $L=1$ level, too, the reduced 
spatial overlap of their orbitals with those of the particles in $L=0$ leads 
to a reduction of the per-particle exchange energy. As the number of particles 
in $L=1$ increases, their intra-level exchange energy also grows and 
overcompensates the initial drop (as seen in the crossing 
of the full and the dashed curve). 

For $g=2$, the initial drop is much more pronounced than for $g<2$, because 
for this value occupation of the spin-down subband of the 
$L=1$ level occurs simultaneously with that of the energetically degenerate 
spin-up subband of the $L=0$ level, so that the exchange energy of some of 
the aditional particles is not only reduced by small spatial overlap but 
strictly zero due to zero spin overlap with the spin-down particles 
already occupying $L=0$. The resulting larger drop may be
so large that for fields and densities compatible with $L\leq 1$, $e_x(n)$ 
does not recover the hypothetical value obtained by filling only the $L=0$ 
level, as illustrated in the upper set of two curves in Fig.~\ref{fig1}.

This complex behaviour of $e_x(n)$ implies that magnetic-field
Thomas-Fermi-Dirac theory,\cite{lai,fushiki,fushiki2} 
current-density-functional theory\cite{vr1,vr2} and 
magnetic-field-density-functional theory\cite{harris1,harris2} 
should have a much richer solution space than
have ordinary Thomas-Fermi-Dirac theory and spin-density-functional
theory, where $e_x(n)$ is strictly monotonous.
The density $n_d$ where the discontinuous drop occurs, and the critical
density $n_c$ where $e_x(n)$ goes through its minimum, depend on the magnetic 
field. For sufficiently large fields
both fall in the metallic density range, {\em i.e.}, both regimes, as well 
as the crossover between them, should be observable in the laboratory.

\section{Higher fields and limiting cases}
\label{specialcases}

In the quantum limit, $B$ is so high that only the $L=0$ level is occupied.
In this limit, the exchange 
energy is given by keeping only $e_x^{(0)}$ in Eq.~(\ref{exLone}). If
the free-electron value $g=2$ is employed, an accidental degeneracy between 
the spin-up subband of $L=0$ level and the spin-down subband of $L=1$ occurs.
Restriction to $L=0$ is thus only rigorously possible if either $g<2$ 
or if the system is fully polarized, so that the spin-up subband of $L=0$ 
level is empty, too. Moreover, the conditions guaranteeing $L=0$
are stricter than those limiting occupation to $L\leq 1$,
given above, and read
\be
\frac{l}{r_sa_0} < \left(\frac{2\sqrt{2}}{3\pi}\right)^{1/3}
\label{l0g2cond}
\ee
for $g=2$, 
\be
\frac{l}{r_sa_0} < \left(\frac{4\sqrt{2}}{3\pi}\right)^{1/3}
\label{l0g0cond}
\ee
for $g=0$, and
\be
\frac{l}{r_sa_0} < 
\left[\frac{2}{3\pi}\left(\sqrt{2-|g|}+\sqrt{2}\right)\right]^{1/3}
\ee
for generic $g\in(-2,2)$.
These conditions are derived in Appendix \ref{appB}.
As long as these conditions are satisfied, Eq.~(\ref{exLone}) with the 
contribution from just $e_x^{(0)}$  can be used for arbitrary occupation 
of the up and down subbands, {\em i.e.} for arbitrary spin polarization.

The further restriction to full spin polarization, {\em i.e.} an empty
spin-up subband, then leads to 
\bea
e_x(n,B)=
\frac{e^2}{8\pi^3}\left(\frac{m \omega_c}{\hbar}\right)^2
\left[C+\ln{\left(p\right)}-e^{p}Ei\left(-p\right)
\right. \nonumber \\ \left.
-p^{1/4}
G^{22}_{23}\left(p\left|\begin{array}{l}
3/4, \, 5/4   \\
3/4, \, 3/4, \, 1/4\\
\end{array}
\right.\right)\right],
\label{dgex3}
\eea
which is the result obtained by DG.\cite{dg71} Here the particle density $n$ 
is equal to the occupation $n_{L=0}^{\sigma=\downarrow}$ and we defined, 
following DG, $p(n,B):=8\pi^4\hbar^3n^2/(m^3\omega_c^3)$. 
As long as no higher levels are occupied and $g=2$, we also have 
$p(n,B)=4 \epsilon_F/ \hbar \omega_c$, where $\epsilon_F$ is the
Fermi energy. (This equality breaks down if other 
spin states or Landau levels are involved, or $g\neq 2$.)

If $\epsilon_F/\hbar \omega_c\ll 1$, then only the bottom of the
lowest subband is occupied, and the preceding equation reduces to
\be
e_x(n,B)=\frac{\pi e^2\hbar}{m \omega_c}n^2\left[\ln(p)-3+C\right].
\label{dgex1}
\ee
In units in which $\hbar=c=1$, Eq.~(\ref{dgex1}) can be written as
\be
e_x(n,B)=\frac{2 \pi e^2}{e B} n^2 \left[\ln \left(
\frac{n}{(eB)^{3/2}}\right)+2.11788 \right],
\label{dgcompare}
\ee
which was first obtained by DG\cite{dg71} and can be directly compared with 
the corresponding equation (47) of BCR,\cite{bcr74}
\be
e_x(n,B)^{BCR}=\frac{2 \pi e^2}{e B} n^2 \left[\ln \left(
\frac{n}{(eB)^{3/2}}\right) + 2.32918 \right],
\label{bcrex}
\ee
featuring a different numerical value inside the brackets. This difference
was also noticed in Refs.~\onlinecite{yonei,fushiki}.

\begin{figure}
\includegraphics[height=60mm,width=75mm,angle=0]{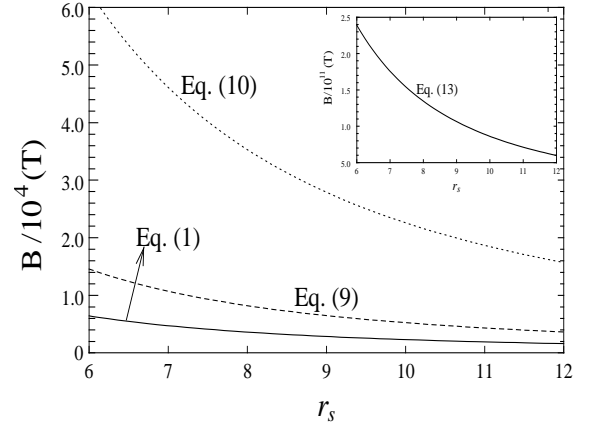}
\caption{\label{fig2}
Applicability of the four expressions discussed here. The curves represent the
{\em lowest} magnetic fields for which the indicated expression is valid, for
the free-electron value of $m$ and $g=2$.}
\end{figure}

For ultra-strong magnetic fields (quantified in Fig.~\ref{fig2}) or for 
extremely low densities, that value, be it $2.11788$ or $2.32918$, becomes 
negligible relative to the logarithmic term, and the DG and BCR expressions 
become identical, both reducing to
\be
\frac{e_x(r_s,B)}{Ry} = -\frac{27}{4\pi}\frac{1}{a_0^3 r_s^6}
\left(\frac{\hbar \omega_c}{Ry}\right)^{-1}
\ln\left(0.141 r_s^2 \frac{\hbar \omega_c}{Ry} \right),
\label{dgex2}
\ee
where $e_x/Ry$ and $\hbar \omega_c /Ry$ denote exchange energy and cyclotron 
energy measured in Rydberg [$1 Ry=e^2/2a_0 = 13.6 eV$]. In DG this limit 
appears as their Eq. (1.1),
\be
\frac{e_x(r_s,B)^{DG}}{Ry} = -\frac{27}{16\pi}\frac{1}{a_0^3 r_s^6}
\left(\frac{\hbar \omega_c}{Ry}\right)^{-1}
\ln\left(0.282 r_s^2 \frac{\hbar \omega_c}{Ry} \right),
\label{dgex2wrong}
\ee
which differs in two ways from Eq.~(\ref{dgex2}):
the numerical factor inside the logarithm is $0.282 = 2 \times 0.141$
and the prefactor is $27/16$, instead of $27/4$.

Our general expression (\ref{exLone}) contains all these limits as special 
cases, and we can therefore verify which of the conflicting expressions is
correct. In the high-field limit of the quantum limit, we find that the DG
Eq.~(\ref{dgcompare}) is correct, while the BCR Eq.~(\ref{bcrex}) is not.
In the ultra-strong field limit, we find that Eq.~(\ref{dgex2}) is correct, 
while the DG Eq.~(\ref{dgex2wrong}) is not. In this latter case, we suspect 
that DG inadvertedly used Hartree units instead of Rydberg units 
($1 H = e^2/ a_0 = 2 Ry$ is the atomic unit of energy), but denoted them as 
Rydberg, as this would explain both the factor of $2$
inside and the factor of $1/4$ outside the logarithm.\cite{footnote1}

Figure~\ref{fig2} illustrates the magnetic field and density regimes for
which each of the above equations is valid. The curves represent the
{\em lowest} magnetic fields for which the indicated expression is valid, for
the free-electron value of $m$ and $g=2$. For all $B(r_s)$ above the full
curve, $L\leq1$ and Eq.~(\ref{exLone}) with arbitrary values of $n_0^\da$, 
$n_0^\ua$, and $n_1^\da$ may be applied. For $g=2$, $n_1^\ua$ must be zero 
to avoid degeneracy with the $L=2$ Landau level, not included in our formula. 
For $g<2$, $n_1^\ua$ is also arbitrary. 

For values of $B(r_s)$ above the dashed curve, Eq.~(\ref{dgex3}) with 
arbitrary values of $n=n_0^\da$ may be applied. For $g=2$, $n_0^\ua$ must be 
zero to avoid degeneracy with the $L=1$ Landau level, not included in the DG
formula. For $g<2$, $n_0^\ua$ is also arbitrary. The relevant expression in
this case is our Eq.~(\ref{exLone}), but with only the $e_x^{(0)}$ term kept.

The restrictions $L=0$ and $L\leq1$ are precisely defined and easily applied.
The restriction $p\ll1$ which leads from Eq.~(\ref{dgex3}) to Eq.~(\ref{dgex1})
and the condition $\left|\ln \left( n/(eB)^{3/2}\right)\right| \gg 2.11788$, 
which leads from Eq.~(\ref{dgex1}) to Eq.~(\ref{dgex2}), are less precisely 
defined, and we simply 
adopt as validity criterium that $p\leq 0.05$ and that $2.11788$ is less than 
5\% of the logarithmic term. From these criteria, we find that for $B(r_s)$ 
above the dotted curve, the high-field limit (\ref{dgex1}) becomes valid. 
The inset shows the values of $B(r_s)$ above which the ultra-high-field limit 
(\ref{dgex2}) is valid. 

To generate Figs.~\ref{fig1} and \ref{fig2} we used the free-electron value 
of the electron mass $m$, requiring very high magnetic fields to satisfy 
conditions (\ref{l1g2cond}) to (\ref{l0g0cond}).
Such high fields are not without physical relevance.
Continuous fields $\sim 40 T$ and pulsed fields $\sim 10^4 T$ can be
produced in the laboratory, white dwarfs have surface fields of order
$\sim 10^4 T$, and neutron stars can have fields in excess of $10^8-10^9 T$.
For effective masses smaller than the free-electron mass, the required fields
are considerably lower: the use of effective masses $m^*=\gamma m$ rescales 
all $B$ values by $B \to \gamma^2 B$, so that a reduction of $m$ by a factor 
of $10$ allows a reduction of $B$ by a factor of $100$. Small effective masses 
and low densities thus bring these fields into the laboratory range.

\section{Conclusions}
\label{concl}

Our key result is Eq.~(\ref{exLone}), for the exchange energy of the 
three-dimensional Fermi gas in magnetic fields for which both the lowest
and the second-lowest Landau level contribute, and the fermions may have
arbitrary spin polarization. This equation predicts the existence of two
physically distinct regimes, one (entered with a discontinuous derivative) 
in which occupation of the $L=1$ level lowers the per-particle exchange energy,
and one (entered through a gradual crossover) in which it increases it. For 
high, but achievable, fields both the discontinuous drop and the gradual 
crossover between the two regimes occur in the metallic density range, so both 
regimes should be experimentally observable. We predict that similar separation
in two regimes occurs every time a new Landau level is included in the 
calculation.

The availability of an analytical expression for the exchange energy of the 
electron gas in high magnetic fields facilitates the construction of density
functionals for magnetic-field Thomas-Fermi-Dirac 
theory,\cite{lai,fushiki,fushiki2}
current-density-functional theory,\cite{vr1,vr2} and 
magnetic-field density-functional theory,\cite{harris1,harris2} but the 
intricate structure of $e_x(n,B)$ implies that their local approximations 
must display a much more complex behaviour than those of ordinary 
Thomas-Fermi-Dirac and spin-density-functional theory.
\\

This work was supported by FAPESP and CNPq. We thank G. Vignale for useful 
remarks on an earlier version of this manuscript.

\appendix
\section{Details on the derivation of Eqs.~(\ref{exLone}) to (\ref{exLone4})} 
\label{appA}

From Eq. (2.1) of DG, and following their procedure until their Eq. (2.5),
we find the per volume exchange energy
\bea
e_x=-\frac{e^2}{4\pi^3}\left(\frac{m\omega_c}{\hbar}\right)^2\sum_{\sigma}
\int_{-\infty}^{\infty}dr_z\int_0^{\infty}d\overline{r}\frac{\overline{r}
e^{-\frac{m\omega_c}{2\hbar}\overline{r}^2}}  
{\left(\overline{r}^2+r_z^2\right)^{1/2}r_z^2}\nonumber \\
\times\left\{\sum_{L=0}^{\infty}\Theta\left(C_L^{\sigma}\right)\,\sin{\left(r_zD_L^{\sigma}\right)}\,
L_L\left(\frac{m\omega_c}{2\hbar}\overline{r}^2\right)\right\}^2 \  \  
\label{exformal}
\eea
where $\Theta(x)$ is the step function, $L_L(x)$ are Laguerre polynomials, 
and we defined
\be
C_L^{\sigma}=\mu-\left(L+\frac{1}{2}\right)\hbar\omega_c-g\mu_0B\sigma,
\ee
\be
D_L^{\sigma}=\sqrt{\frac{2m}{\hbar^2}\left(\mu-
\left(L+\frac{1}{2}\right)\hbar\omega_c-g\mu_0B\sigma\right)},
\ee
where $\mu_0=e\hbar/2mc$ is the Bohr magneton.

Keeping the contributions from the $L=0$ and $L=1$ levels and performing
the change of variable 
\be
\bar{r}^2=\frac{2\hbar}{m\omega_c}\left(r_zt+\frac{\hbar}{2m\omega_c}t^2\right)
\ee
we obtain
\begin{widetext}
\bea
e_x&=&-\frac{e^2}{2\pi^3}\frac{m\omega_c}{\hbar}\sum_{\sigma}\int_0^{\infty}dt\,
e^{-\frac{\hbar
t^2}{2m\omega_c}}\int_0^{\infty}dr_z\,\frac{e^{-r_zt}}{r_z^2}\left\{\Theta(C_0^{\sigma})
\sin^2{(D_0^{\sigma}r_z)} \right. \nonumber \\
&+&\left.\Theta(C_1^{\sigma})\sin^2{(D_1^{\sigma}r_z)}\left(1-tr_z-\frac{\hbar
t^2}{2m\omega_c}\right)^2\right. \nonumber \\
&+& \left. 
2\,\Theta(C_0^{\sigma})\Theta(C_1^{\sigma})\sin{(D_0^{\sigma}r_z)}\sin{(D_1^{\sigma}r_z)}
\left(1-tr_z-\frac{\hbar t^2}{2m\omega_c}\right)
\right\}
\label{exintermed1}
\eea

Using Laplace Transforms to perform the $r_z$ integral,\cite{erdely} we find
\bea
e_x=-\frac{e^2}{2\pi^3}\frac{m\omega_c}{\hbar}\sum_{\sigma}\int_0^{\infty}dt\,
e^{-\frac{\hbar t^2}{2m\omega_c}}
\left\{\Theta(C_0^{\sigma})\left[D_0^{\sigma}\tan^{-1}{\left(\frac{2D_0^{\sigma}}{t}\right)}
-\frac{1}{4}t\ln{\left(1+\frac{4{D_0^{\sigma}}^2}{t^2}\right)}\right]  \right. \nonumber\\
\left. +\Theta(C_0^{\sigma})\Theta(C_1^{\sigma})\left[\left(t-\frac{\hbar}{4m\omega_c}t^3\right)
\ln{\left(\frac{t^2+\left(D_0^{\sigma}-D_1^{\sigma}\right)^2}{t^2+\left(D_0^{\sigma}+D_1^{\sigma}\right)^2}\right)} \right. \right.\nonumber \\
\left.\left.+\left(\frac{\hbar}{2m\omega_c}t^2-1\right)\left(D_0^{\sigma}-D_1^{\sigma}\right)\tan^{-1}{\left(\frac{D_0^{\sigma}-D_1^{\sigma}}{t}\right)} \right.\right. \nonumber\\
\left. \left. 
+\left(1-\frac{\hbar}{2m\omega_c}t^2\right)\left(D_0^{\sigma}+D_1^{\sigma}\right)\tan^{-1}{\left(\frac{D_0^{\sigma}+D_1^{\sigma}}{t}\right)}\right] \right. \nonumber \\
\left. +\Theta(C_1^{\sigma})\left[D_1^{\sigma}\left(1-\frac{\hbar}{2m\omega_c}t^2\right)^2
\tan^{-1}{\left(\frac{2D_1^{\sigma}}{t}\right)}\right.\right. \nonumber\\
\left. \left. 
+\left(-\frac{3t}{4}+\frac{\hbar}{2m\omega_c}t^3-\frac{\hbar^2}{16m^2\omega_c^2}t^5\right)
\ln{\left(1+\frac{4{D_1^{\sigma}}^2}{t^2}\right)}+2{D_1^{\sigma}}^2\frac{t}{t^2+4{D_1^{\sigma}}^2}\right]\right\}
\label{exintermed2}
\eea
\end{widetext}
The integral over the first term, involving only contributions from $L=0$, can 
be calculated following the procedure of Appendix A of the DG paper.\cite{dg71}
The other integrals are calculated by integration by parts and Laplace 
transforms. In particular, the integrals involving $\tan^{-1}{(x)}$ can be 
calculated by using the result of the first integral.

$C_L^{\sigma}$ and $D_L^{\sigma}$ can be written in terms of the occupation 
number $n_L^{\sigma}$ as $C_L^{\sigma}=(2\pi^4\hbar^2l^4/m) {n_L^{\sigma}}^2$ 
and $D_L^{\sigma}=2\pi^2l^2n_L^{\sigma}$. The step functions turn out to be
unnecessary, because the term arrising from $L=1$, and the mixed term 
containing contributions from $L=0$ and $L=1$, are automatically zero when
$n_1^{\sigma}=0$. We thus find the exchange energy as given Eqs.
(\ref{exLone}) - (\ref{exLone4}). More details of the calculation are 
available through Ref.~\onlinecite{jmphd}.

An extension of these analytical calculations to higher $L$ values seems
extraordinarily cumbersome. To include, {\em e.g.}, $L=2$ one must keep one
more term in the sum of Eq.~(\ref{exformal}), which gives rise to not just 
one more intra-level integral in Eq.~(\ref{exintermed1}), but also to many 
new inter-level integrals. In the general case, for arbitrary $L$, there is no
guarantee that the resulting integrals can all be solved by Laplace transforms,
and reduced to some known special function. The three-dimensional case is, in 
this regard, more complicated than the two-dimensional one,\cite{2dex} because 
of the additional $k_z$ dependence of the single-particle energies and the 
resulting Landau-level dispersion. Numerical work on the general case is 
under way.

\section{Details on the derivation of Eqs.~(\ref{l1g2cond}) to (\ref{l0g0cond})}
\label{appB}

The conditions specifying the fields and densities for which restriction
to $L=0$ or $L\leq1$ is valid depend on the Landau-level degeneracy. The 
single-particle dispersion in the spin $\sigma$ sublevel of Landau level $L$ 
is
\be
\epsilon(k_z,L,\sigma)=\frac{\hbar^2}{2m}k_z^2+\left(L+\frac{1}{2}\right)\hbar \omega_c+g\mu_0B\sigma.
\label{dispersion}
\ee
In each subband, the Fermi momentum is related to the partial density by
$k_{FL}^{\sigma}=2\pi^2l^2n_L^{\sigma}$. The degeneracy of these levels 
depends on the value of $g$.
For the free-electron value $g=2$, the spin-up subband of level $L$ and the 
spin-down subband of level $L+1$ are degenerate. For $g=0$, on the other hand,
the spin-up and spin-down sublevels of Landau level $L$ are degenerate. 
We first deal with $g=2$.

The restriction to $L=0$ implies 
\bea
\epsilon(k_z=0,L=1, \downarrow)=\epsilon(k_z=0,L=0, \uparrow) 
\\
> \epsilon_F =\epsilon(k_{F0}^\da,L=0,\downarrow).
\eea
From Eq. (\ref{dispersion}), and using the fact that 
for $L=0$ and $g=2$, $n_0^\da=n$, we immediately find
\be
n^2<\frac{1}{2\pi^4l^6},
\ee
which is equivalent to Eq. (\ref{l0g2cond}) of the main text. Similarly,
the restriction to $L\leq 1$ implies 
\bea
\epsilon(k_z=0,L=2, \downarrow)=\epsilon(k_z=0,L=1, \uparrow)
\\ >\epsilon_F=\epsilon(k_{F1}^\da,1,\da)=\epsilon(k_{F0}^\ua,0,\ua)
=\epsilon(k_{F0}^\da,0,\da).
\eea
The total density is $n=n_0^{\downarrow}+n_0^{\uparrow}+n_1^{\downarrow}$. 
Solving this set of equations for $n$ yields
\be
n < \frac{2+\sqrt{2}}{\sqrt{2}\pi^2l^3},
\ee
which is equivalent to Eq. (\ref{l1g2cond}) of the main text.

The corresponding conditions for $g=0$ follow in the same way. The case $g=0$
is occasionally used as a methodological device
in theoretical work, because it allows to cleanly separate the effects of spin 
magnetism from those of orbital magnetism. Experimentally, $g=0$ occurs, 
{\em e.g.}, in systems studied in the context of $g$-factor engineering of 
devices.\cite{gfaceng} 

When $|g| < 2$ and $g\neq 0$, no such degeneracy occurs. The restriction to 
$L=0$ then implies
\be
n < \frac{\sqrt{2-|g|}+\sqrt{2}}{2\pi^2l^3},
\ee
where $n=n_0^{\uparrow}+n_0^{\downarrow}$.

%]})
\end{document}